\documentclass[a4paper,12pt]{article}

\usepackage[latin1]{inputenc}
\usepackage[T1]{fontenc}
\usepackage[english]{babel}

\usepackage[margin=25mm]{geometry}

\usepackage{amsmath,amssymb}

\newcommand*{\bs}{\boldsymbol}

\newcommand*{\mbf}{\mathbf}
\newcommand*{\mcal}{\mathcal}
\newcommand*{\mrm}{\mathrm}

\newcommand*{\diff}{\mathop{}\!\mathrm{d}}
\newcommand*{\pd}{\partial}
\newcommand*{\im}{\ensuremath{\mathrm{i}}}
\newcommand*{\e}{\mathop{\mathrm{e}}\nolimits}
\newcommand*{\const}{\ensuremath{\mathrm{const}}}
\newcommand*{\R}{\mathbb{R}}
\newcommand*{\T}{\mathrm{T}}

\newcommand*{\aver}[1]{\langle#1\rangle}

\newcommand*{\Four}{\mathcal{F}}
\newcommand*{\Lapl}{\mathcal{L}}
\newcommand*{\HE}{\mathrm{HE}}
\newcommand*{\BC}{\mathrm{BC}}

\newcommand*{\disc}{\mathrm{disc}}
\newcommand*{\reg}{\mathrm{reg}}
\newcommand*{\cont}{\mathrm{cont}}
\newcommand*{\sing}{\mathrm{sing}}

\newcommand*{\kB}{k_{\mathrm{B}}^{}}

\numberwithin{equation}{section}
\allowdisplaybreaks[4]

\usepackage[numbers, sort&compress]{natbib}

\usepackage[nottoc]{tocbibind}

\newcommand*{\ie}{i.\,e.}
\newcommand*{\eg}{e.\,g.}

\usepackage{graphicx}
\usepackage[font=small,labelfont=bf]{caption}
\usepackage{xcolor}

\usepackage[unicode,colorlinks]{hyperref}

\definecolor{darkblue}{rgb}{0,0,0.8}

\hypersetup{%
  bookmarks=true,
  bookmarksopen=true,
  bookmarksopenlevel=2,
  bookmarksnumbered=true,
  linkcolor=darkblue,
  citecolor=darkblue,
  urlcolor=darkblue,
  pdfpagelayout=OneColumn,
  pdfstartview={XYZ null null 1},
  pdfview={XYZ null null null}
}

\begin{document}

\title{%
  \bfseries
  Hyperbolicity of the ballistic-conductive model of heat conduction: the reverse side of the coin
}

\author{%
  \bfseries
  S.~A.~Rukolaine\thanks{E-mail address: \texttt{rukol@ammp.ioffe.ru}}
}

\date{%
  \small
  Ioffe Institute, 26 Polytekhnicheskaya, St.\,Petersburg 194021, Russia
}

\maketitle

\begin{abstract}
  The heat equation, based on Fourier's law, is  commonly used for description of heat conduction. However, Fourier's law is valid under the assumption of local thermodynamic equilibrium, which is violated in very small dimensions and short timescales, and at low temperatures. In the paper R.~Kov\'acs and P.~V\'an, Generalized heat conduction in heat pulse experiments, \emph{Int. J. Heat Mass Transf.}, 83:613--620, 2015, a ballistic-conductive (BC) model of heat conduction was developed. In this paper we study the behavior of solutions to an initial value problem (IVP) in 1D in the framework of the linearized ballistic-conductive (BC) model. As a result of the study, an unphysical effect has been found when part of the initial thermal energy does not spread anywhere.
\end{abstract}

\section{Introduction}
\label{sec:Intro}

The heat equation, based on Fourier's law, is commonly used for description of heat conduction. However, Fourier's law is valid under the assumption of local thermodynamic equilibrium \cite{deGrootMazur:1984, KondepudiPrigogine:2015}, which is violated in very small dimensions and short timescales, and at low temperatures~\cite{JosephPreziosi:1989, JosephPreziosi:1990, DreyerStruchtrup:1993, GuoWang:2015, JouCimmelli:2016, BothEtAl:2016, Zhang:2020, Chen:2021, Zhmakin:2021}.

The first (and maybe most ``famous'') modification of Fourier's law was proposed by Cattaneo \cite{Cattaneo:1948} and later and independently by Vernotte \cite{Vernotte:1958} (see a historical review in Refs.~\cite{JosephPreziosi:1989, JosephPreziosi:1990}). Cattaneo's equation leads to a so called hyperbolic heat equation (HHE) (or telegraph equation, though this is not quite correct, since there is a difference between these equations). However, in a number of works, it has been shown that the hyperbolic heat equation has significant flaws in describing heat conduction, see, \eg, Refs.~\cite{PorraEtAl:1997, KornerBergmann:1998, ShiomiMaruyama:2006, BrightZhang:2009, ZhangEtAl:2011, Auriault:2016, RukolaineChistiakova:2016, Zhang:2020}%
\footnote{Refs.~\cite{PorraEtAl:1997, RukolaineChistiakova:2016} discuss mass transfer. However, the obtained unphysical solutions of the telegraph equation in 2D and 3D, when they may take negative values, even if the initial values are non-negative, apply to the hyperbolic heat equation as well.}.
The papers~\cite{BrightZhang:2009, Maillet:2019} critically discuss experimental ``validations'' of the HHE.

There are a number of methods for constructing models of non-Fourier heat conduction.
Guyer and Krumhansl \cite{GuyerKrumhansl:1964, GuyerKrumhansl:1966a, GuyerKrumhansl:1966b} solved the linearized phonon Boltzmann equation and obtained an equation which described second sound in solids. However, the Guyer-Krumhansl equation fails in describing ballistic phonons.
Dreyer and Struchtrup \cite{DreyerStruchtrup:1993} used a simplified model of the phonon Boltzmann equation in Callaway's approximation~\cite{Callaway:1959} to derive an infinite system of coupled moment equations. The simplification was that all phonons travelled with the Debye velocity. To close the system, \ie, to reduce it to a finite system, Dreyer and Struchtrup used the maximum entropy principle. This resulted in a finite symmetric hyperbolic system of partial differential equations for the moments. A shortcoming of this approach is that a small number of moment equations, amenable to solution, gives an incorrect speed of propagation of ballistic phonons. To obtain a suitable propagation speed, quite a large number of equations must be taken.
Mohammadzadeh and Struchtrup~\cite{MohammadzadehStruchtrup:2017} used the Callaway model with frequency-dependent relaxation time to derive macroscopic moment equations for phonon heat transport at room temperature. However, to achieve satisfactory results, many moments are needed, which makes it problematic to model heat transfer with so many equations. Besides, phonon polarization was not taken into account in Ref.~\cite{MohammadzadehStruchtrup:2017}.

Tzou proposed the dual-phase-lag (DPL) model~\cite{Tzou:1997, TzouXu:2011} in which a constitutive equation is a modification of Fourier's law with two delay times. 
While the DPL model \emph{per se} is senseless \cite{JordanEtAl:2008, DreherEtAl:2009, OrdonezMirandaAlvaradoGil:2011, Rukolaine:2014}, its derivatives are questionable from the physical point of view \cite{FabrizioFranchi:2014jts, Rukolaine:2014, Rukolaine:2017}.

The framework of rational extended thermodynamics (RET) \cite{MullerRuggeri:1998, RuggeriSugiyama:2015} consists of a hierarchy of balance laws, similar to the infinite system of coupled moment equations in the kinetic theory of rarefied gases. The closure procedure uses the entropy principle, and principle of causality and stability. Shortcomings of RET are similar to those of Dreyer and Struchtrup's approach.

Extended irreversible thermodynamics (EIT) \cite{JouEtAl:2010} uses dissipative fluxes as independent non-equilibrium state variables. Constitutive equations in EIT are derived on the basis of the balance laws with the use of the second law.
The simplest constitutive equation derived in the framework of EIT is Cattaneo's equation.

General Equation for the Non-Equilibrium Reversible-Irreversible Coupling (GENE\-RIC) \cite{Ottinger:2005, PavelkaEtAl:2018} determines the dynamics of state variables through the sum of reversible and irreversible contributions. The main problem in the framework of GENERIC is to find matrices-operators determining the contributions.

The thermomass theory~\cite{DongEtAl:2011} uses equivalence between thermal energy and mass, based on Einstein's mass-energy relation, and a hydrodynamic-like model of heat conduction was constructed in the framework of this theory.
A phenomenological hybrid phonon gas (HPG) model, taking into account both longitudinal and transverse phonons, was developed in Refs.~\cite{Ma:2013jht, Ma:2013ijhmt}.

Non-equilibrium thermodynamics with internal variables (NET-IV) \cite{MauginMuschik:1994PartI, MauginMuschik:1994PartII, VanFulop:2012} is similar to EIT but more general and flexible: instead of dissipative fluxes it uses tensor state variables, which in general have no direct physical meaning.
In this case such variables are considered as fitting parameters. However, the advantage of the internal variables is that they provide thermodynamically consistent models.
A comparison of RET, GENERIC and NET-IV is given in Ref.~\cite{OttingerEtAl:2020, KovacsEtAl:2021cmt, SzucsEtAl:2022}.

In Ref.~\cite{KovacsVan:2015} a ballistic-conductive (BC) model of heat conduction in the framework of NET-IV was developed. In Refs.~\cite{KovacsVan:2016, Kovacs:2017Thesis, KovacsVan:2018} a linearized form of this model was tested on experimental data and showed a quantitative description of heat transfer by transversal ballistic phonons, while demonstrating only qualitative description of the second sound. At the same time, the ballistic signal of longitudinal phonons was not described at all. The linearized form of the BC model is described by a symmetric hyperbolic system of partial differential equations. This ensures the finite speed of heat propagation.

Note that a system of equations similar to the system of the linearized BC model, but with different coefficients, was obtained in the framework of the kinetic theory of phonons \cite{DreyerStruchtrup:1993, MullerRuggeri:1998}. A similar model was also obtained in the framework of EIT~\cite{NettletonSobolev:1996}.

The linearized BC model is of interest because it can be considered not only in the framework of nonequilibruim thermodynamics, but also as a hyperbolic approximation to the phonon Boltzmann equation, which is also hyperbolic. From this point of view, the Cattaneo model (hyperbolic heat equation) is the first hyperbolic approximation to the phonon Boltzmann equation, while the linearized BC model is the second one (see similar approximations in the kinetic theory of phonons~\cite{DreyerStruchtrup:1993, MullerRuggeri:1998}). Both models provide finiteness of the heat propagation velocity. However, this is not enough for the model to accurately describe heat transfer, as the example of the hyperbolic heat equation shows (see the criticism of the hyperbolic heat equation in Ref.~\cite{Zhang:2020}). Since the linearized BC model was tested on experimental data, its theoretical study is of considerable interest.

In this paper we study the behavior of solutions to an initial value problem (IVP) in 1D in the framework of the linearized BC model. Analysis of the solutions to the IVP, rather than to the initial boundary value problem with the same source term and initial conditions, allows us to establish the properties of the model itself without the influence of boundary conditions.
However, since the speed of heat propagation in the BC model is finite, adding any boundary conditions would not alter the conclusions.
This paper is a sequel of the conference paper~\cite{Rukolaine:2023SPbPUJ}.

The paper is organized as follows.
In Section~\ref{sec:StatementIVP} we formulate the initial value problem.
In Section~\ref{sec:SolutionIVP} we solve the initial value problem.
Finally, in Section~\ref{sec:Concl} we provide concluding remarks.

\section{Statement of the initial value problem}
\label{sec:StatementIVP}

We consider the system of dimensionless equations in 1D, describing the linearized form of the ballistic-conductive model of heat conduction \cite{KovacsVan:2016, Kovacs:2017Thesis, KovacsVan:2018}:
\begin{subequations}
  \label{eq:BCModel}
  \begin{align}
    \label{eq:BCModelA}
    &T_t^{}
      + q_x^{}
      + \gamma T
      =
      f(x,t)
    \\[1ex]
    \label{eq:BCModelB}
    &\tau_q^{} q_t^{}
      + q
      + T_x^{}
      + \kappa Q_x^{}
      = 0
    \\[1ex]
    \label{eq:BCModelC}
    &\tau_Q^{} Q_t^{}
      + Q
      + \kappa q_x^{}
      = 0
  \end{align}
\end{subequations}
or, equivalently,
\begin{equation*}
  \bs{w}_t^{}
  + \mcal{A} \bs{w}_x^{}
  + \mcal{B} \bs{w}
  =
  \begin{pmatrix}
    f & 0 & 0
  \end{pmatrix}^\T,
\end{equation*}
where $^\T$ means transposition,
\begin{equation}
  \label{eq:Matrices}
  \bs{w}
  =
  \begin{pmatrix}
    T \\ q \\[0.5ex] Q
  \end{pmatrix},
  \quad
  \mcal{A}
  =
  \begin{pmatrix}
    0 & 1 & 0
    \\[1ex]
    1/\tau_q^{} & 0 & \kappa/\tau_q^{}
    \\[1ex]
    0 & \kappa/\tau_Q^{} & 0
  \end{pmatrix},
  \quad
  \mcal{B}
  =
  \begin{pmatrix}
    \gamma & 0 & 0
    \\[1ex]
    0 & 1/\tau_q^{} & 0
    \\[1ex]
    0 & 0 & 1/\tau_Q^{}
  \end{pmatrix},
\end{equation}
$T$ is temperature\footnote{In Refs.~\cite{KovacsVan:2015, KovacsVan:2016, Kovacs:2017Thesis, KovacsVan:2018} the dimensionless temperature $T$ is multiplied by a coefficient $\tau_{\Delta}^{}$. Its presence or absence does not affect the conclusions of this paper.}, $q$ is a heat flux, $f$ is a heat source, $Q$ is an internal variable, $\gamma$ is a heat exchange parameter or volumetric heat transfer coefficient (as it is called in Refs.~\cite{KovacsVan:2016, Kovacs:2017Thesis, KovacsVan:2018}), $\tau_q^{}$ and $\tau_Q^{}$ are relaxation times, and $\kappa$ is a dissipation parameter.
Note that the designations for the dimensionless values and parameters differ from those in Refs.~\cite{KovacsVan:2015, KovacsVan:2016, Kovacs:2017Thesis, KovacsVan:2018}.
Eq.~\eqref{eq:BCModelA} is the standard energy balance equation, including term responsible for heat exchange.
Eqs.~\eqref{eq:BCModelB} and \eqref{eq:BCModelC} are constitutive ones, obtained in the framework of NET-IV.
The internal variable $Q$ is similar to a second-order dissipative flux in the framework of EIT. The relaxation times $\tau_q^{}$ and $\tau_Q^{}$ are the characteristic times for which the variables $q$ and $Q$ return to equilibrium after excitation, respectively.

If $\kappa=0$, then Eq.~\eqref{eq:BCModelB} takes the form of Cattaneo's equation. If $\tau_Q^{} = 0$, then the system~\eqref{eq:BCModel} leads to the Guyer-Krumhansl equation. If $\tau_q^{} = 0$ and $\kappa=0$, then Eq.~\eqref{eq:BCModelB} takes the form of Fourier's law. Thus, the linearized BC model generalizes the well-known models.

The system~\eqref{eq:BCModel} is hyperbolic~\cite{CourantHilbert:1962}, and the eigenvalues of the matrix $\mcal{A}$, Eq.~\eqref{eq:Matrices}, are
\begin{equation}
  \label{eq:Eigenvalues}
  \lambda_{1,2}^{} = \pm v,
  \quad
  v
  =
  \sqrt{\frac{1}{\tau_q^{}} \left( 1 + \frac{\kappa^2}{\tau_Q^{}} \right)},
  \quad\text{and}\quad
  \lambda_3^{} = 0.
\end{equation}
Hyperbolicity of the system~\eqref{eq:BCModel} ensures the finite speed of heat propagation equal to $v$, since the corresponding characteristics are given by the equations $x \pm v t = \const$. However, in addition to the non-zero eigenvalues of the matrix $\mcal{A}$, there is the zero eigenvalue, and the corresponding characteristic is given by the equation $x = \const$. This has an important consequence, as will be shown below.

Note that a system of equations similar to the system~\eqref{eq:BCModel}, but with different coefficients, was obtained in the framework of the kinetic theory of phonons \cite{DreyerStruchtrup:1993, MullerRuggeri:1998}. However, the system in Refs.~\cite{DreyerStruchtrup:1993, MullerRuggeri:1998} results in an incorrect propagation speed.

Excluding in the system \eqref{eq:BCModel} the variables $q$ and $Q$, we find that temperature satisfies the equation
\begin{multline}
  \label{eq:TEq}
  \tau_q^{} \tau_Q^{} T_{ttt}^{}
  + \left( \tau_q^{} + \tau_Q^{} + \gamma \tau_q^{} \tau_Q^{} \right) T_{tt}^{}
  + \left[ 1 + \gamma \left( \tau_q^{} + \tau_Q^{} \right) \right] T_t^{}
  + \gamma T
  - \left( 1 + \gamma \kappa^2 \right) T_{xx}^{}
  \\[1ex]
  - \left( \tau_Q^{} + \kappa^2 \right) T_{xxt}^{}
  =
  \tau_q^{} \tau_Q^{} f_{tt}^{}
  + \left( \tau_q^{} + \tau_Q^{} \right) f_t^{}
  + f
  - \kappa^2 f_{xx}^{},
  \quad
  x \in \R,
\end{multline}
and we consider the equation on the entire real axis. If $\gamma = 0$ and $f=0$, Eq.~\eqref{eq:TEq} takes the form of Eq.~(14) in Ref.~\cite{KovacsVan:2016}.
Note that an equation, similar to Eq.~\eqref{eq:TEq} (with $\gamma = 0$), was obtained earlier in Ref.~\cite{NettletonSobolev:1996} in the framework of EIT.

We assume that the medium was at rest until the moment $t=0$, and the initial temperature distribution was equal to zero.
Therefore, we impose zero initial conditions on the system~\eqref{eq:BCModel}:
\begin{equation}
  \label{eq:InitCondBCModel}
  T \rvert_{t=0}^{}
  = 0,
  \quad
  q \rvert_{t=0}^{}
  = 0,
  \quad
  Q \rvert_{t=0}^{}
  = 0.
\end{equation}
Substituting these conditions into Eqs.~\eqref{eq:BCModelB} and \eqref{eq:BCModelC}, we obtain the conditions
\begin{equation}
  \label{eq:pdtqQ}
  q_t^{} \rvert_{t=0}^{}
  = 0,
  \qquad
  Q_t^{} \rvert_{t=0}^{}
  = 0.
\end{equation}
As a result, we obtain the initial conditions for Eq.~\eqref{eq:TEq}:
\begin{equation}
  \label{eq:ICforTEq}
  T \rvert_{t=0}^{}
  = 0,
  \quad
  T_t^{} \rvert_{t=0}^{}
  =
  f \rvert_{t=0}^{},
  \quad
  T_{tt}^{} \rvert_{t=0}^{}
  =
  \left( f_t^{} - \gamma f \right) \rvert_{t=0}^{}
\end{equation}
(the second condition is obtained from Eq.~\eqref{eq:BCModelA} and conditions~\eqref{eq:InitCondBCModel}, the third condition follows from Eqs.~\eqref{eq:BCModelA}, \eqref{eq:BCModelB} and conditions~\eqref{eq:pdtqQ}).

\section{Solution of the initial value problem}
\label{sec:SolutionIVP}

We consider Eq.~\eqref{eq:TEq} on the entire real axis.
In order to reveal the features of the BC model, we assume that at the moment $t=0$ a finite amount of thermal energy was released at the origin, namely,
\begin{equation}
  \label{eq:FDeltaDelta}
  f(x,t)
  =
  \delta(x) \delta(t),
\end{equation}
where $\delta(\cdot)$ is the Dirac delta-function. The solution to the problem~\eqref{eq:TEq}, \eqref{eq:ICforTEq} with the source term~\eqref{eq:FDeltaDelta} is the fundamental solution, and we denote it by $\varPhi$.

The solution to the problem~\eqref{eq:TEq}, \eqref{eq:ICforTEq} with a general heat source term $f$ is given by
\begin{equation*}
  T(x,t)
  =
  \frac{1}{2\pi} \int_{-\infty}^\infty \Four T(\xi, t) \e^{-\im \xi x} \diff \xi,
\end{equation*}
where $\Four T$ is the Fourier transform.
The Fourier transform of the fundamental solution is given by (see Appendix~\ref{sec:FourTransfOfSolutionInstantF})
\begin{equation}
  \label{eq:FourFundamentalSolution}
  \Four \varPhi(\xi,t)
  =
  \e^{-\mu_1^{}(\xi) t} E(\xi)
  + \e^{-\mu_2^{}(\xi) t}
  \left[
    F(\xi) \cos B(\xi) t
    + G(\xi) \frac{\sin B(\xi) t}{B(\xi)}
  \right],
\end{equation}
where the coefficients $\mu_1^{}$, $\mu_2^{}$ are given by Eq.~\eqref{eq:MuOneTwo}, and the other coefficients are given by Eqs.~\eqref{eq:ABCD}, \eqref{eq:EFG}.

It follows from Eq.~\eqref{eq:FourFundamentalSolution} that that the fundamental solution can be represented in the form (see Appendix~\ref{sec:FundamentalSolution})
\begin{equation}
  \label{eq:TDelta}
  \varPhi(x,t)
  =
  \begin{cases}
    \varPhi_{\sing,1}^{}(x,t)
    + \varPhi_{\sing,2}^{}(x,t)
    + \varPhi_\reg^{}(x,t),
    & \lvert x \rvert \leq v t,
    \\[1ex]
    0,
    & \lvert x \rvert > v t,
  \end{cases}
\end{equation}
where
\begin{equation}
  \label{eq:TSingOne}
  \varPhi_{\sing,1}^{}(x,t)
  =
  \e^{-\mu_{1,\infty}^{} t} E_\infty^{} \delta(x)
\end{equation}
and
\begin{equation}
  \label{eq:TSingTwo}
  \varPhi_{\sing,2}^{}(x,t)
  =
  \e^{-\mu_{2,\infty}^{} t}
  F_\infty^{} \frac{1}{2} \left[
    \delta \left( x - v t \right)
    + \delta \left( x + v t \right)
  \right]
\end{equation}
are the singular parts of the fundamental solution, while $\varPhi_\reg^{}$ is its regular part.
The coefficients $\mu_{1,\infty}^{}$, $\mu_{2,\infty}^{}$ and $E_\infty^{}$, $F_\infty^{}$ are given by Eqs.~\eqref{eq:MuInfty} and \eqref{eq:EFGInfty}, respectively.
It is important to emphasize that Eq.~\eqref{eq:TDelta} is just a mathematical representation of the fundamental solution. This does not contradict the non-additivity of temperature.

One can conclude from the representation~\eqref{eq:TDelta} that
\begin{equation}
  \label{eq:LimSingular}
  \lim_{\varepsilon \to 0}^{} \int_{-\varepsilon}^\varepsilon \varPhi(x,t) \diff x
  =
  \e^{-\mu_{1,\infty}^{} t} E_\infty^{}.
\end{equation}
This means that part of the initial thermal energy does not spread anywhere, though this part decreases exponentially with time.
This effect is due to the term $\varPhi_{\sing,1}^{}$. The mathematical reason for this effect is the zero eigenvalue of the matrix $\mcal{A}$, see Eq.~\eqref{eq:Eigenvalues}. This effect of the BC model appears to be unphysical. Indeed, in dielectric crystals, heat conduction is due to phonons.
Phonons are bosons, and in the equilibrium state the average number of phonons as a function of frequency satisfies the Planck distribution (see, \eg, \cite{Kittel:2005, Kaviany:2014})
\begin{equation}
  \label{eq:PlanckDistr}
  \aver{n_\omega^{}}
  =
  \frac{1}{\exp\!\left( \dfrac{\hbar \omega}{\kB T} \right) - 1},
\end{equation}
where $\omega$ is the phonon frequency, $\hbar$ is the reduced Planck constant, $\kB$ is the Boltzmann constant. According to the Planck distribution, at low temperatures only low-frequency acoustic phonons contrubute to heat transfer. These are ballistic phonons, \ie, acoustic phonons that have a maximum propagation velocity close to the speed of sound. The group velocity of phonons is given by $v_{\mrm{g}}^{} = \pd\omega(k)/\pd k$, where $\omega(k)$ is the dispersion relation, and $k$ is the wave number. Zero-velocity phonons are either higher-frequency acoustic phonons or optical phonons, which are also high-frequency (see, \eg, \cite{Kaviany:2014}). In a weakly nonequilibrium state, the contribution of zero-velocity phonons is negligible as well. This means that the singular part $\varPhi_{\sing,1}^{}$ of the fundamental solution must be negligible.

One can also conclude from the representation~\eqref{eq:TDelta} that
\begin{equation}
  \label{eq:LimBallistic}
  \lim_{\varepsilon \to 0}^{} \int_{vt - \varepsilon}^{vt + \varepsilon} \varPhi(x,t) \diff x
  =
  \lim_{\varepsilon \to 0}^{} \int_{-vt - \varepsilon}^{-vt + \varepsilon} \varPhi(x,t) \diff x
  =
  \frac{1}{2} \e^{-\mu_{2,\infty}^{} t} F_\infty^{}.
\end{equation}
This means that equal parts of the initial thermal energy propagate in opposite directions at the speed $v$, while exponentially decaying. This is due to the term $\varPhi_{\sing,2}^{}$, which corresponds to ballistic phonons.

\begin{figure}[!htb]
  \centering
  \includegraphics*[width=\linewidth]{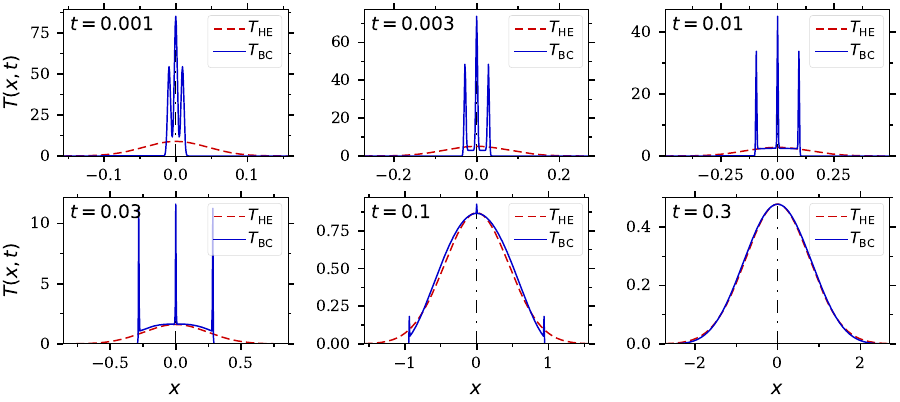}
  \caption{%
    The solution $T_\BC^{}$ to the problem~\eqref{eq:TEq}, \eqref{eq:ICforTEq} with the instantaneous heat source~\eqref{eq:FPhiDelta} in the framework of the ballistic-conductive model in comparison with the solution $T_\HE^{}$, Eq.~\eqref{eq:THeatEq}, to the similar problem~\eqref{eq:HeatEqHomo}, \eqref{eq:ICHeatEqHomo} for the heat equation. The dimensionless parameters are $\tau_q^{} = 0.0202$, $\tau_Q^{} = 0.0077$ and $\kappa = 0.078$ ($\kappa^2 \approx 0.0061$), $\gamma = 0.250$. The asymptotic values are $E_\infty^{} \approx 0.44$, $\mu_{1,\infty}^{} \approx 73$, $F_\infty^{} \approx 0.56$, $\mu_{2,\infty}^{} \approx 53$. Thermal energy contained in the unphysical part of the solution is comparable to that in the ``ballistic'' part (see Eqs.~\eqref{eq:LimSingular} and~\eqref{eq:LimBallistic}).}
  \label{fig:BC_eq_IVP_T=11K}
\end{figure}

Figs.~\ref{fig:BC_eq_IVP_T=11K}--{}\ref{fig:BC_eq_IVP_T=145K} present the solution to the problem~\eqref{eq:TEq}, \eqref{eq:ICforTEq} with the instantaneous heat source $f$, obtained for various values of the dimensionless parameters $\tau_q^{}$, $\tau_Q^{}$ and $\kappa$. The dimensional quantities have been taken from Refs.~\cite{KovacsVan:2015, KovacsVan:2016, Kovacs:2017Thesis, KovacsVan:2018}, see Appendix~\ref{sec:Parameters}.
To make the solution more visual, instead of delta-like initial distribution $\delta(x)$ in Eq.~\eqref{eq:FDeltaDelta}, we have taken the function
\begin{equation*}
  \varphi_\sigma^{}(x)
  =
  \frac{1}{\sqrt{2 \pi} \sigma} \exp \left( -\frac{x^2}{2 \sigma^2} \right),
\end{equation*}
where $\sigma = 0.002$. In this case the heat source is given by
\begin{equation}
  \label{eq:FPhiDelta}
  f(x,t)
  =
  \varphi_\sigma^{}(x) \delta(t)
\end{equation}
instead of Eq.~\eqref{eq:FDeltaDelta}.
Therefore, instead of the fundamental solution, the figures present the close, but more visual, solution
\begin{equation*}
  T(x,t)
  =
  \int_{-\infty}^\infty \varPhi(x-y,t) \varphi_\sigma^{}(y) \diff y,
\end{equation*}
which is the convolution of the fundamental solution with $\varphi_\sigma^{}$ (see Eq.~\eqref{eq:FourTFInstant}).
This solution shows the singular parts in a more visual form, while practically not distorting the regular part.
The figures clearly show the unphysical part~\eqref{eq:TSingOne} (the immovable peak in the middle) and the ``ballistic'' part \eqref{eq:TSingTwo} (two peaks running in opposite directions).
The figures show that thermal energy contained in the unphysical part of the solution is not negligible, on the contrary, it is comparable or even greater than in the ``ballistic'' part.
This is confirmed by the values of $E_\infty^{}$, $\mu_{1,\infty}^{}$ and  $F_\infty^{}$, $\mu_{2,\infty}^{}$ (see Eqs.~\eqref{eq:LimSingular} and~\eqref{eq:LimBallistic}) shown in the figure captions.

The solution to the problem~\eqref{eq:TEq}, \eqref{eq:ICforTEq} with the heat source~\eqref{eq:FPhiDelta} is compared in the figures with the solution to the heat equation
(with the same heat source)
\begin{equation}
  \label{eq:HeatEqHomo}
  T_t^{}
  - T_{xx}^{}
  + \gamma T
  =
  \varphi_\sigma^{}(x) \delta(t)
\end{equation}
with the zero initial condition
\begin{equation}
  \label{eq:ICHeatEqHomo}
  T \rvert_{t=0}^{}
  = 0.
\end{equation}
The solution to the problem~\eqref{eq:HeatEqHomo}, \eqref{eq:ICHeatEqHomo} is given by
\begin{equation}
  \label{eq:THeatEq}
  T(x,t)
  =
  \frac{1}{2 \sqrt{\pi t}} \int_{-\infty}^\infty \exp\left( -\frac{(x-y)^2}{4t} - \gamma t \right) \varphi_\sigma^{}(y) \diff y.
\end{equation}
If $\varphi_\sigma^{}(x) = \delta(x)$, then
\begin{equation*}
  T(x,t)
  =
  \frac{1}{2 \sqrt{\pi t}} \exp\left( -\frac{x^2}{4t} - \gamma t \right).
\end{equation*}

\begin{figure}[!htb]
  \centering
  \includegraphics*[width=\linewidth]{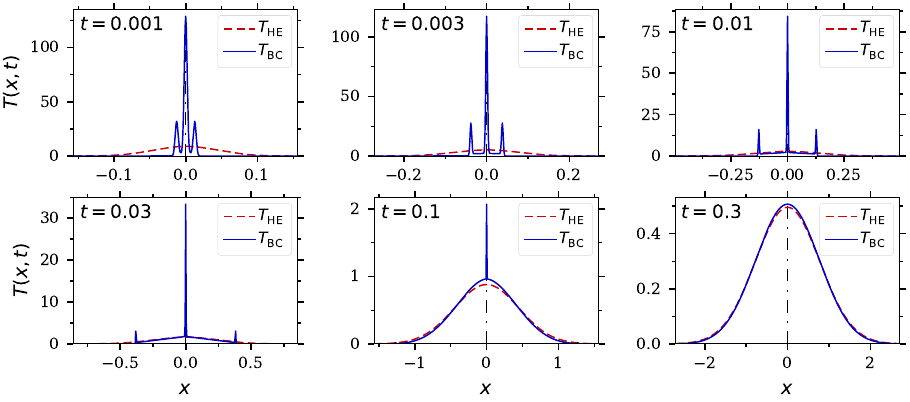}
  \caption{%
    The solutions to the same problems as in Fig.~\ref{fig:BC_eq_IVP_T=11K}.
    The dimensionless parameters are $\tau_q^{} = 0.0186$, $\tau_Q^{} = 0.0070$ and $\kappa = 0.118$ ($\kappa^2 \approx 0.014$), $\gamma = 0.130$. The asymptotic values are $E_\infty^{} \approx 0.67$, $\mu_{1,\infty}^{} \approx 48$, $F_\infty^{} \approx 0.33$, $\mu_{2,\infty}^{} \approx 74$. The unphysical effect is more pronounced than in Fig.~\ref{fig:BC_eq_IVP_T=11K}. Thermal energy contained in the unphysical part of the solution is significantly greater than that in the ``ballistic'' part (see Eqs.~\eqref{eq:LimSingular} and~\eqref{eq:LimBallistic}).}
  \label{fig:BC_eq_IVP_T=13K}
\end{figure}

\begin{figure}[!htb]
  \centering
  \includegraphics*[width=\linewidth]{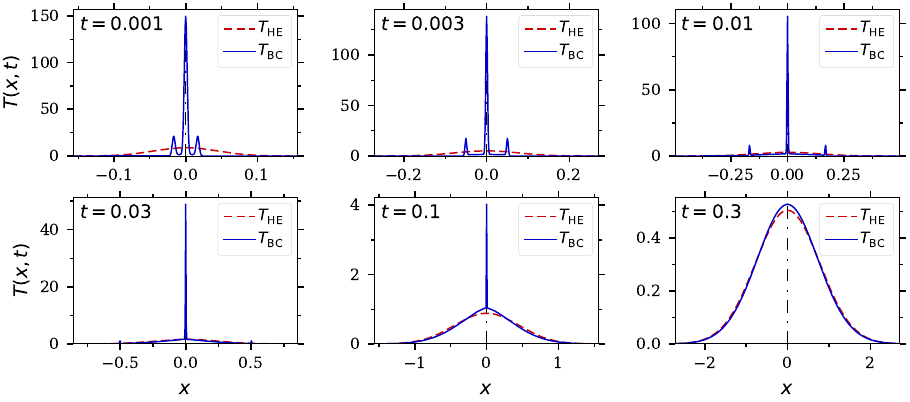}
  \caption{%
    The solutions to the same problems as in Fig.~\ref{fig:BC_eq_IVP_T=11K}.
    The dimensionless parameters are $\tau_q^{} = 0.0156$, $\tau_Q^{} = 0.0058$ and $\kappa = 0.140$ ($\kappa^2 \approx 0.020$), $\gamma = 0.076$. The asymptotic values are $E_\infty^{} \approx 0.77$, $\mu_{1,\infty}^{} \approx 39$, $F_\infty^{} \approx 0.23$, $\mu_{2,\infty}^{} \approx 99$. The unphysical effect is more pronounced than in Figs.~\ref{fig:BC_eq_IVP_T=11K} and~\ref{fig:BC_eq_IVP_T=13K}. Thermal energy contained in the unphysical part of the solution is significantly greater than that in the ``ballistic'' part (see Eqs.~\eqref{eq:LimSingular} and~\eqref{eq:LimBallistic}).}
  \label{fig:BC_eq_IVP_T=145K}
\end{figure}

\section{Conclusions}
\label{sec:Concl}

In this paper we have studied the initial value problem (IVP) for Eq.~\eqref{eq:TEq}, describing heat transfer in the framework of the ballistic-conductive model~\eqref{eq:BCModel}.
Analysis of the solution to the IVP has allowed us to establish its structure and properties and, therefore, to identify the features of the model itself without the influence of boundary conditions.
The matrix $\mcal{A}$ (see Eq.~\eqref{eq:Matrices}) of the system~\eqref{eq:BCModel} has three eigenvalues (see Eq.~\eqref{eq:Eigenvalues}). Two of the eigenvalues are equal to $\pm v$, and this ensures the finite speed of heat propagation equal to $v$. This is due to the characteristics given by the equations $x \pm v t = \const$. However, the third eigenvalue is equal to zero, which corresponds to the characteristic given by the equation $x = \const$. This has an important consequence, namely that in the BC model part of the initial thermal energy does not spread anywhere.

To determine how strongly this effect can be manifested, we have examined the BC model with parameters obtained from an experiment. The study has shown that thermal energy contained in the immovable part of the solution (see Eq.~\eqref{eq:LimSingular}) can be significantly greater than that in the ``ballistic'' part (see Eq.~\eqref{eq:LimBallistic}).
The ratio remains valid at any scale.
Since this effect is observed in the fundamental solution, it will also be the case for any source term.

This behavior of the model is unphysical.
Indeed, in dielectric crystals, heat conduction is due to phonons. Phonons are bosons, and in the equilibrium state the average number of phonons as a function of frequency satisfies the Planck distribution, see Eq.~\eqref{eq:PlanckDistr}. According to the Planck distribution, at low temperatures only low-frequency acoustic phonons contribute to heat transfer. These are ballistic phonons, \ie{}, phonons that have a maximum propagation velocity close to the speed of sound. Zero-velocity phonons are either higher-frequency acoustic phonons or optical phonons, which are also high-frequency ones. In a weakly non-equilibrium state, the contribution of the zero-velocity phonons is negligible as well.
Calculations have shown that thermal energy contained in the term, corresponding to zero-velocity phonons (see Eq.~\eqref{eq:LimSingular}), is not negligible, on the contrary, it is comparable or even significantly greater than in the term, corresponding to ballistic phonons (see Eq.~\eqref{eq:LimBallistic}), and Figs.~\ref{fig:BC_eq_IVP_T=11K}--\ref{fig:BC_eq_IVP_T=145K} show this.
This means that the effect (that thermal energy contained in zero-velocity phonons greater than in the ballistic phonons) is manifested for  any heat source, \ie{}, thermal energy is distributed between phonons in exactly the same way, regardless of how close the system is to the equilibrium state or far from it.
The same conclusions are valid in the case of a sample of finite thickness due to the finite velocity of heat propagation.

Note that a similar effect in a mass transfer model, described by the Jeffreys-type equation, was found in Ref.~\cite{RukolaineSamsonov:2013}, although this equation is not hyperbolic.

The results obtained in this paper can serve as an addition to the discussion on hyperbolic models in thermodynamics given in Ref.~\cite{SzucsEtAl:2020}. The advantage of hyperbolic models is considered to be the finiteness of the heat propagation velocity. However, hyperbolicity can also have a downside, as can be seen from the results of this paper.

Besides, the question of approximation of the phonon Boltzmann equation remains open. Solving the Boltzmann equation is a time-consuming computational problem. Therefore, models of not very high order that would approximate the Boltzmann equation are of considerable interest. Since the Boltzmann equation is hyperbolic, hyperbolic models may seem preferable. However, it turns out that low-order hyperbolic models (the hyperbolic heat equation and the BC model) have serious flaws in describing heat conduction, and therefore cannot be such approximations.

Note that similar problems arise in the field of radiative heat transfer \cite{Modest:2013, HowellEtAl:2021}. It is described by the radiative transfer equation (RTE), which is the linear Boltzmann equation. Approximations to the RTE face problems similar to those that arise when approximating the phonon Boltzmann equation. Various hyperbolic approximations to the RTE exhibit unphysical behavior of various kinds \cite{McClarrenEtAl:2008JCP, SchaferEtAl:2011MMS, Modest:2013}.

\appendix

\section{The Fourier transform of the solution to the initial value problem}

\subsection{General heat source $f$}
\label{sec:FourTransfOfSolutionGeneralF}

The Fourier transform of the equation~\eqref{eq:TEq} and the initial conditions \eqref{eq:ICforTEq} results in the equation
\begin{multline}
  \label{eq:FourKovacsVanEq}
  \tau_q^{} \tau_Q^{} \Four T_{ttt}^{}
  + \left( \tau_q^{} + \tau_Q^{} + \gamma \tau_q^{} \tau_Q^{} \right) \Four T_{tt}^{}
  + \left[ 1 + \gamma \left( \tau_q^{} + \tau_Q^{} \right) + \left( \tau_Q^{} + \kappa^2 \right) \xi^2 \right] \Four T_t^{}
  \\[1ex]
  + \left[ \gamma + \left( 1 + \gamma \kappa^2 \right) \xi^2 \right] \Four T
  =
  \tau_q^{} \tau_Q^{} \Four f_{tt}^{}
  + \left( \tau_q^{} + \tau_Q^{} \right) \Four f_t^{}
  + \Four f
  + \kappa^2 \xi^2 \Four f
\end{multline}
and initial conditions
\begin{equation}
  \label{eq:FourICKovacsVanEq}
  \Four T \rvert_{t=0}^{}
  = 0,
  \quad
  \Four T_t^{} \rvert_{t=0}^{}
  =
  \Four f \rvert_{t=0}^{},
  \quad
  \Four T_{tt}^{} \rvert_{t=0}^{}
  =
  \left( \Four f_t^{}  - \gamma \Four f \right) \rvert_{t=0}^{},
\end{equation}
where the Fourier transform is given by
  $\Four T(\xi, \cdot)
  =
  \int_{-\infty}^\infty T(x, \cdot) \e^{\im \xi x} \diff x$.
The Laplace transform of the equation~\eqref{eq:FourKovacsVanEq} with the initial conditions \eqref{eq:FourICKovacsVanEq} results in the equation
\begin{multline*}
  \big\{
    \tau_q^{} \tau_Q^{} s^3
    + \left( \tau_q^{} + \tau_Q^{} + \gamma \tau_q^{} \tau_Q^{} \right) s^2
    + \left[ 1 + \gamma \left( \tau_q^{} + \tau_Q^{} \right) + \left( \tau_Q^{} + \kappa^2 \right) \xi^2 \right] s
  \\[1ex]
  + \left[ \gamma + \left( 1 + \gamma \kappa^2 \right) \xi^2 \right] \big\} \Lapl\Four T
  =
  \left[
    \tau_q^{} \tau_Q^{} s^2
    + \left( \tau_q^{} + \tau_Q^{} \right) s
    + 1
    + \kappa^2 \xi^2
  \right] \Lapl\Four f,
\end{multline*}
where the Laplace transform is given by
  $\Lapl T(\cdot, s)
  =
  \int_0^\infty T(\cdot, t) \e^{- s t} \diff t$.
Hence, the Laplace-Fourier transform of the solution is given by
\begin{multline}
  \label{eq:LaplFourTGen}
  \Lapl\Four T(\xi,s)
  =
  \\[1ex]
  \frac{%
    [\tau_q^{} \tau_Q^{} s^2
    + (\tau_q^{} + \tau_Q^{}) s
    + 1
    + \kappa^2 \xi^2] \Lapl\Four f(\xi,s)
  }{%
    \tau_q^{} \tau_Q^{} s^3
    + (\tau_q^{} + \tau_Q^{} + \gamma \tau_q^{} \tau_Q^{}) s^2
    + [1 + \gamma (\tau_q^{} + \tau_Q^{}) + (\tau_Q^{} + \kappa^2) \xi^2] s
    + \gamma + \left( 1 + \gamma \kappa^2 \right) \xi^2
  }.
\end{multline}

\subsection{Instantaneous heat source $f = \varphi(x) \delta(t)$}
\label{sec:FourTransfOfSolutionInstantF}

If the heat source term is equal to $f = \varphi(x) \delta(t)$, then its Laplace-Fourier transform is equal to $\Lapl\Four f(\xi,s) = \Four\varphi(\xi)$.

If $\varphi(x) = \delta(x)$, then $\Four\varphi(\xi) = 1$. In this case the solution is the fundamental one, and its Laplace-Fourier transform is given by (see Eq.~\eqref{eq:LaplFourTGen})
\begin{multline}
  \label{eq:LaplFourT}
  \Lapl\Four \varPhi(\xi,s)
  =
  \frac{%
    s^2
    + a_0^{} s
    + (\tau_q^{} \tau_Q^{})^{-1} \left( 1  + \kappa^2 \xi^2 \right)
  }{%
    s^3 + a s^2 + b s + c
  }
  \\[1ex]
  =
  \frac{u^2 + C u + D}{(u - 2 A) \left[ \left( u + A \right)^2 + B^2 \right]}
  =
  \frac{E}{u - 2 A}
  + \frac{F (u + A) + G}{\left( u + A \right)^2 + B^2},
\end{multline}
where
\begin{multline}
  \label{eq:abc}
  a
  =
  a_0^{} + \gamma,
  \quad
  b(\xi)
  =
  \frac{1}{\tau_q^{} \tau_Q^{}}
  + \gamma a_0^{} + v^2 \xi^2,
  \quad
  c(\xi)
  =
  \frac{\gamma + (1 + \gamma \kappa^2) \xi^2}{\tau_q^{} \tau_Q^{}},
  \\[1ex]
  \quad
  a_0^{}
  =
  \frac{1}{\tau_q^{}} + \frac{1}{\tau_Q^{}},
\end{multline}
$v$ is given by Eq.~\eqref{eq:Eigenvalues}, $s^3 + a s^2 + b s + c = u^3 + \chi u + \psi$,
\begin{equation}
  \label{eq:u}
  s
  =
  u - \frac{a}{3},
\end{equation}
\begin{equation}
  \label{eq:pq}
  \chi(\xi)
  =
  - \frac{a^2}{3}
  + b,
  \qquad
  \psi(\xi)
  =
  2 \left( \frac{a}{3} \right)^3
  - \frac{a}{3} b
  + c,
\end{equation}
\begin{multline}
  \label{eq:ABCD}
  A(\xi)
  =
  \frac{\alpha + \beta}{2},
  \quad
  B(\xi)
  =
  \sqrt{3} \,\frac{\alpha - \beta}{2},
  \quad
  C
  =
  \frac{a_0^{} - 2\gamma}{3},
  \\[1ex]
  \quad
  D(\xi)
  =
  \frac{-2 a_0^2 - \gamma a_0^{} + \gamma^2}{9}
  + \frac{1 + \kappa^2 \xi^2}{\tau_q^{} \tau_Q^{}},
\end{multline}
\begin{equation}
  \label{eq:AlphaBetaDelta}
  \alpha(\xi)
  =
  \sqrt[3]{-\frac{\psi}{2} + \sqrt{\varDelta}},
  \qquad
  \beta(\xi)
  =
  \sqrt[3]{-\frac{\psi}{2} - \sqrt{\varDelta}},
  \qquad
  \varDelta(\xi)
  =
  \left( \frac{\chi}{3} \right)^3
  + \left( \frac{\psi}{2} \right)^2,
\end{equation}
the roots $\alpha$ and $\beta$ are chosen so that the equality $\alpha \beta = - \chi/3$ is valid and the value $A$ is real,
and
\begin{multline}
  \label{eq:EFG}
  E(\xi)
  =
  \frac{4 A^2 + 2 A C + D}{9 A^2 + B^2},
  \quad
  F(\xi)
  =
  1 - E(\xi),
  \\[1ex]
  G(\xi)
  =
  \frac{-3 A^3 + A B^2 + (3 A^2 + B^2) C - 3 A D}{9 A^2 + B^2}.
\end{multline}
Taking into account Eq.~\eqref{eq:u}, one can conclude that the inverse Laplace transform is equal to
\begin{equation*}
  \Lapl^{-1}
  \left[
    \frac{E}{u - 2 A}
    + \frac{F (u + A) + G}{\left( u + A \right)^2 + B^2}
  \right]
  =
  \e^{-\mu_1^{} t} E
  + \e^{-\mu_2^{} t}
  \left[
    F \cos B t
    + G \frac{\sin B t}{B}
  \right],
\end{equation*}
where
\begin{equation}
  \label{eq:MuOneTwo}
  \mu_1^{} \equiv \mu_1^{}(\xi)
  =
  - 2A
  + \frac{a}{3},
  \quad
  \mu_2^{} \equiv \mu_2^{}(\xi)
  =
  A
  + \frac{a}{3}.
\end{equation}
As a result, we obtain from Eq.~\eqref{eq:LaplFourT} that the Fourier transform of the fundamental solution is given by Eq.~\eqref{eq:FourFundamentalSolution}.

In the case of a general distribution $\varphi$ the Fourier transform of the solution $T$ is given by
\begin{equation}
  \label{eq:FourTFInstant}
  \Four T(\xi,t)
  =
  \Four \varPhi(\xi,t) \Four\varphi(\xi).
\end{equation}

\section{Asymptotic behavior of the coefficients as $\xi \to \infty$}

It follows from Eqs.~\eqref{eq:pq} and \eqref{eq:abc} that the coefficients $\chi$ and $\psi$ are given by
\begin{equation*}
  \chi(\xi)
  =
  v^2 \xi^2 + c'
  \quad\text{and}\quad
  \psi(\xi)
  =
  C_\psi^{} \xi^2 + c'',
\end{equation*}
where $v$ is given by Eq.~\eqref{eq:Eigenvalues},
\begin{equation}
  \label{eq:CpCq}
  C_\psi^{}
  =
  \frac{1 + \gamma \kappa^2}{\tau_q^{} \tau_Q^{}}
  - \frac{1}{3} \left( \frac{1}{\tau_q^{}} + \frac{1}{\tau_Q^{}} + \gamma \right) v^2,
\end{equation}
the expressions for the coefficients $c'$ and $c''$ are not significant.
Hence the asymptotic behavior of $\varDelta$ (see Eq.~\eqref{eq:AlphaBetaDelta}) is given by
\begin{equation*}
  \varDelta
  =
  \left( \frac{v^2}{3} \right)^3 \xi^6 \left[
    1 + \left( \frac{v^4 c'}{9} + \frac{C_\psi^2}{4} \right) \left( \frac{v^2}{3} \right)^{-3} \frac{1}{\xi^2}
    + O\left( \frac{1}{\xi^4} \right)
  \right],
\end{equation*}
and, therefore,
\begin{equation*}
  \sqrt{\varDelta}
  =
  \left( \frac{v^2}{3} \right)^{3/2} \lvert \xi \rvert^3 \left[
    1 + \frac{1}{2} \left( \frac{v^4 c'}{9} + \frac{C_\psi^2}{4} \right) \left( \frac{v^2}{3} \right)^{-3} \frac{1}{\xi^2}
    + O\left( \frac{1}{\xi^4} \right)
  \right].
\end{equation*}
This results in the asymptotic behavior of the radicals $\alpha$ and $\beta$ (Eq.~\eqref{eq:AlphaBetaDelta})
\begin{equation*}
  \alpha(\xi)
  =
  \left( \frac{v^2}{3} \right)^{1/2} \lvert \xi \rvert \left[
    1 - \left( \frac{v^2}{3} \right)^{-3/2} \frac{C_\psi^{}}{6} \frac{1}{\lvert \xi \rvert}
    + \frac{K}{\xi^2}
    + O \left( \frac{1}{\lvert \xi \rvert^3} \right)
  \right]
\end{equation*}
and
\begin{equation*}
  \beta(\xi)
  =
  - \left( \frac{v^2}{3} \right)^{1/2} \lvert \xi \rvert \left[
    1 + \left( \frac{v^2}{3} \right)^{-3/2} \frac{C_\psi^{}}{6} \frac{1}{\lvert \xi \rvert}
    + \frac{K}{\xi^2}
    + O \left( \frac{1}{\lvert \xi \rvert^3} \right)
  \right],
\end{equation*}
where the expression for the coefficient $K$ is not significant.

The asymptotic behavior of the coefficients $A$, $B$ and $D$ (Eq.~\eqref{eq:ABCD}) follows from the above asymptotics and is given by
\begin{multline*}
  A
  =
  - \frac{C_\psi^{}}{2 v^2}
  + O \left( \frac{1}{\xi^2} \right)
  \equiv
  \frac{1}{6} \left( \frac{1}{\tau_q^{}} + \frac{1}{\tau_Q^{}} + \gamma \right)
  - \frac{1 + \gamma \kappa^2}{2 (\tau_Q^{} + \kappa^2)}
  + O \left( \frac{1}{\xi^2} \right),
  \\[1ex]
  \quad
  B
  =
  v \lvert \xi \rvert \left[ 1 + O\left( \frac{1}{\xi^2} \right) \right]
  \quad\text{and}\quad
  D
  =
  \frac{\kappa^2}{\tau_q^{} \tau_Q^{}} \xi^2 \left[ 1 + O\left( \frac{1}{\xi^2} \right) \right].
\end{multline*}
Therefore, the asymptotic behavior of the coefficients $E$, $F$ and $G$ (Eq.~\eqref{eq:EFG}) is given by
\begin{equation}
  \label{eq:EFGAsy}
  E
  =
  E_\infty^{} + O \left( \frac{1}{\xi^2} \right),
  \quad
  F
  =
  F_\infty^{} + O \left( \frac{1}{\xi^2} \right)
  \quad\text{and}\quad
  G
  =
  G_\infty^{} + O \left( \frac{1}{\xi^2} \right),
\end{equation}
where
\begin{equation}
  \label{eq:EFGInfty}
  E_\infty^{}
  =
  \frac{\kappa^2}{\tau_Q^{} + \kappa^2},
  \quad
  F_\infty^{}
  =
  1 - E_\infty^{},
\end{equation}
and the expression for the coefficient $G_\infty^{}$ is unimportant.

The asymptotic behavior of the coefficients $\mu_1^{}$ and $\mu_2^{}$ (Eq.~\eqref{eq:MuOneTwo}) is given by
\begin{equation*}
  \mu_1^{}(\xi)
  =
  \mu_{1,\infty}^{}
  + O \left( \frac{1}{\xi^2} \right),
  \qquad
  \mu_2^{}(\xi)
  =
  \mu_{2,\infty}^{}
  + O \left( \frac{1}{\xi^2} \right).
\end{equation*}
where
\begin{equation}
  \label{eq:MuInfty}
  \mu_{1,\infty}^{}
  =
  \frac{1 + \gamma \kappa^2}{\tau_Q^{} + \kappa^2},
  \qquad
  \mu_{2,\infty}^{}
  =
  \frac{1}{2} \left(
    \frac{1}{\tau_q^{}}
    + \frac{1}{\tau_Q^{}}
    + \gamma
    - \frac{1 + \gamma \kappa^2}{\tau_Q^{} + \kappa^2}
  \right).
\end{equation}

As a result, we obtain the asymptotics
\begin{equation}
  \label{eq:Asymptotics}
  \begin{gathered}
    \e^{-\mu_1^{} t} E
    =
    \e^{-\mu_{1,\infty}^{} t} E_\infty^{}
    + O \left( \frac{1}{\xi^2} \right)
    \quad\text{as}\quad
    \xi \to \infty,
    \\
    \e^{-\mu_2^{} t} F \cos B t
    =
    \e^{-\mu_{2,\infty}^{} t} F_\infty^{} \cos v \xi t
    + O \left( \frac{1}{\xi^2} \right)
    \quad\text{as}\quad
    \xi \to \infty,
    \\
    \e^{-\mu_2^{} t} G \frac{\sin B t}{B}
    =
    \e^{-\mu_{2,\infty}^{} t} G_\infty^{} \frac{\sin v \xi t}{v \xi}
    + O \left( \frac{1}{\xi^2} \right)
    \quad\text{as}\quad
    \xi \to \infty.
  \end{gathered}
\end{equation}

\section{The fundamental solution}
\label{sec:FundamentalSolution}

Taking into account the asymptotics~\eqref{eq:Asymptotics}, we obtain the asymptotic behavior of the Fourier transform
\begin{equation*}
  \Four \varPhi
  =
  \e^{-\mu_{1,\infty}^{} t} E_\infty^{}
  + \e^{-\mu_{2,\infty}^{} t}
  \left[
    F_\infty^{} \cos v \xi t
    + G_\infty^{} \frac{\sin v \xi t}{v \xi}
  \right]
  + O \left( \frac{1}{\xi^2} \right)
  \quad\text{as}\quad
  \xi \to \infty,
\end{equation*}
where the coefficients are given by Eqs.~\eqref{eq:EFGInfty} and \eqref{eq:MuInfty}.
Note, that
\begin{equation*}
  \Four^{-1} 1
  =
  \delta(x),
  \qquad
  \Four^{-1} \left[ \cos \!\left( v \xi t \right) \right]
  =
  \frac{1}{2} \left[
    \delta \left( x - v t \right)
    + \delta \left( x + v t \right)
  \right],
\end{equation*}
\begin{equation*}
  \Four^{-1} \left[ \frac{1}{\xi} \sin \!\left( v \xi t \right) \right]
  =
  \frac{1}{2} \mbf{1}_{(-v t, \,v t)}^{},
\end{equation*}
where $\Four^{-1}$ is the inverse Fourier transform,
and
\begin{equation}
  \label{eq:Indicator}
  \mbf{1}_S^{}(x)
  =
  \begin{cases}
    1, & x \in S,
    \\
    0, & x \notin S,
  \end{cases}
\end{equation}
is the indicator function of a set $S$.
Therefore, one can conclude that the fundamental solution can be represented in the form
\begin{equation*}
  \varPhi(x,t)
  =
  \varPhi_{\sing,1}^{}(x,t)
  + \varPhi_{\sing,2}^{}(x,t)
  + \varPhi_\disc^{}(x,t)
  + \varPhi_\cont^{}(x,t),
\end{equation*}
where $\varPhi_{\sing,1}^{}$ and $\varPhi_{\sing,2}^{}$ are singular parts given by
\begin{equation*}
  \varPhi_{\sing,1}^{}(x,t)
  =
  \e^{-\mu_{1,\infty}^{} t} E_\infty^{} \delta(x)
\end{equation*}
and
\begin{equation*}
  \varPhi_{\sing,2}^{}(x,t)
  =
  \e^{-\mu_{2,\infty}^{} t}
  F_\infty^{} \frac{1}{2} \left[
    \delta \left( x - v t \right)
    + \delta \left( x + v t \right)
  \right],
\end{equation*}
$\varPhi_\disc^{}$ is a discontinuous part given by
\begin{equation*}
  \varPhi_\disc^{}(x,t)
  =
  \e^{-\mu_{2,\infty}^{} t} G_\infty^{} \frac{1}{2 v} \mbf{1}_{(-v t, \,v t)}^{},
\end{equation*}
$\varPhi_\cont^{}(x,t)$ is a continuous function, since its Fourier transform has the asymptotic behavior $O(1/\xi^2)$ as $\xi \to \infty$ \cite{Zorich2:2004}.

Note, that the system~\eqref{eq:BCModel} is hyperbolic and the maximum absolute value of the eigenvalues is equal to $v$. Therefore, perturbations propagate at the speed $v$. Since the initial disturbance is at the origin, one can conclude that $\varPhi(x,t) = 0$ for $\lvert x \rvert > v t$, and hence $\varPhi_\cont^{}(x,t) = 0$ for $\lvert x \rvert > v t$ as well.
Finally, we define the regular part by $\varPhi_\reg^{} = \varPhi_\disc^{} + \varPhi_\cont^{}$, which is equal to zero for $\lvert x \rvert > v t$ and continuous for $\lvert x \rvert < v t$. As a result, the solution $\varPhi$ takes the form of Eq.~\eqref{eq:TDelta}.

\section{The values of the dimensionless parameters}
\label{sec:Parameters}

In this section, we denote dimensional quantities with an asterisk, as in Refs.~\cite{Ma:2013jht, Ma:2013ijhmt}.
The quantities have been taken from Refs.~\cite{KovacsVan:2015, KovacsVan:2016, Kovacs:2017Thesis, KovacsVan:2018}.

The speed of sound (in a transverse phonon gas) is $v^* = 3186 \,\text{m/s}$.
The length of a sample is $L^* = 7.9 \,\text{mm}$.
The heat pulse duration is $\Delta t^* = 0.24\,\mu\text{s}$.
The mass density is $\rho = 2866 \,\text{kg/m$^3$}$.
The thermal diffusivity is given by
\begin{equation*}
  \alpha
  =
  \frac{k}{\rho c},
\end{equation*}
where $k$, $\rho$ and $c$ are the thermal conductivity, mass density and specific heat, respectively.
The dimensional propagation speed is given by \cite{Kovacs:2017Thesis}
\begin{equation*}
  v^*
  =
  \sqrt{\frac{1}{\tau_q^*} \left( \alpha + \frac{{\kappa^*}^2}{\tau_Q^*} \right)}.
\end{equation*}
Therefore, the dimensional dissipation parameter $\kappa^*$ can be calculated from this formula.

Dimensionless values and parameters are given by \cite{KovacsVan:2015, KovacsVan:2016, Kovacs:2017Thesis, KovacsVan:2018}
\begin{equation*}
  x
  =
  \frac{x^*}{L},
  \quad
  t
  =
  \frac{\alpha t^*}{L^2},
  \quad
  \tau_{q,Q}^{}
  =
  \frac{\alpha \tau_{q,Q}^*}{L^2},
  \quad
  \kappa
  =
  \frac{\kappa^*}{L},
  \quad
  \gamma
  =
  \frac{\Delta t^*}{\rho c} \gamma^*,
  \quad
  v
  =
  \frac{L v^*}{\alpha}.
\end{equation*}

Dimensional parameters for the sample at $11$\,K are $k = 8573 \,\text{W/(m K)}$, $c = 1.118 \,\text{J/(kg K)}$, $\tau_q^* = 0.471\,\mu\text{s}$, $\tau_Q^* = 0.18\,\mu\text{s}$, $\gamma^* = 3.34 \,\text{W/(mm$^3$ K)}$.
Then $\alpha = 2.68 \,\text{m$^2$/s}$ and, therefore, $\tau_q^{} = 0.0202$, $\tau_Q^{} = 0.00772$, $\kappa = 0.0780$, $\gamma = 0.250$, see Fig.~\ref{fig:BC_eq_IVP_T=11K}.

Dimensional parameters for the sample at $13$\,K are $k = 10{,}200 \,\text{W/(m K)}$, $c = 1.8 \,\text{J/(kg K)}$, $\tau_q^* = 0.586\,\mu\text{s}$, $\tau_Q^* = 0.22\,\mu\text{s}$, $\gamma^* = 2.8 \,\text{W/(mm$^3$ K)}$.
Then $\alpha = 1.98 \,\text{m$^2$/s}$ and, therefore, $\tau_q^{} = 0.0186$, $\tau_Q^{} = 0.0070$, $\kappa = 0.118$, $\gamma = 0.130$, see Fig.~\ref{fig:BC_eq_IVP_T=13K}.

Dimensional parameters for the sample at $14.5$\,K are $k = 10{,}950 \,\text{W/(m K)}$, $c = 2.543 \,\text{J/(kg K)}$, $\tau_q^* = 0.65\,\mu\text{s}$, $\tau_Q^* = 0.24\,\mu\text{s}$, $\gamma^* = 2.31 \,\text{W/(mm$^3$ K)}$.
Then $\alpha = 1.50 \,\text{m$^2$/s}$ and, therefore, $\tau_q^{} = 0.0156$, $\tau_Q^{} = 0.0058$, $\kappa = 0.140$, $\gamma = 0.076$, see Fig.~\ref{fig:BC_eq_IVP_T=145K}.

\renewcommand{\bibname}{References}
\bibliographystyle{unsrt}

\end{document}